# Water on The Moon, II. Origins & Resources

Arlin Crotts (Columbia University)

In Part I we recount the history of observation and laboratory measurement culminating with the excavation of water from a permanently shadowed region (PSR) near the lunar South Pole by the impact of the *LCROSS* mission in 2009. In this installment we consider what the current data imply about the nature of water and other volatile substances on and in the Moon.

*LCROSS* hints of increasing volatile concentration at increasing depths, to 3 meters or more into the Moon, as indicated by amount of water seen as the event developed. [1] In comparison new epithermal neutron results indicate that within the first two meters depth of regolith resides a desiccated layer in the top 0.4 meters. *LRO*'s LEND (Lunar Exploration Neutron Detector) found 4% water-equivalent hydrogen in Cabeus versus 5.6 ± 2.9% dug out by *LCROSS*. [2] These agree, but may indicate more water at greater depth at the impact.

[1] "A Model for the Distribution of Volatiles at the LCROSS Impact Site" by A. Colaprete, J. Heldmann, D. H. Wooden, K. Mjaseth, M. Shirley, W. Marshall, R. Elphic, B. Hermalyn & P. Schultz, 2011, in "*A Wet vs. Dry Moon*" Lunar & Planetary Institute, 2011 June 13-16, #6011.

[2] "Hydrogen Mapping of the Lunar South Pole Using the LRO Neutron Detector Experiment LEND" by I. G. Mitrofanov et al., 2010, *Science*, 330, 483.

Few seem to realize the extent to which LEND found that hydrogen in the lunar soil correlates only modestly with PSRs. Many PSRs (about 70%) contain little hydrogen, and some strongest neutron-absorbing areas (about 80%) are not PSRs but often isolated highlands regions far from deep craters, in the lunar polar region but hundreds of kilometers from the poles. [3] There is no evidence that the non-PSR hydrogen sites mark the impact of a volatile-laden asteroid or comet. It is plausible that the hydrogen in these regions leaked from inside the Moon rather than froze out of the thin lunar atmosphere. We detail this below.

[3] "Which Spot on the Moon has the Highest Content of Hydrogen?" by Anton Sanin on behalf

of LEND Team, 2010, *Lunar Exploration & Analysis Group*, (http://www.lpi.usra.edu/meetings/leag2010/presentations/WedAM/saninEtAl.pdf); and "Relationship between Hydrogen-enriched Areas and Permanently Shadowed Regions near the Lunar South Pole" by William Boynton, Jerry Droege, Igor Mitrofanov & LEND team, 2010, *Lunar Exploration & Analysis Group*, (http://www.lpi.usra.edu/meetings/leag2010/presentations/WedAM/boyntonEtAl.pdf): despite its weakness, the PSR-hydrogen correlation (25% of area) is stronger than random (6%), but this is tied almost entirely to craters Shoemaker and Cabeus.

We imagine various origins of lunar water: delivery by comets and asteroids, endogenous sources (fire fountains, outgassing, apatite), solar wind implantation, Solar System dust, or even interstellar clouds. The amount of hydroxyl on the lunar surface is considerable, roughly 10 million tonnes at any instant, and much more over the Moon's history. [4] OH/$H_2O$ inside the Moon is also huge, estimated as 40 trillion to quadrillions of tonnes; [5] with presumably only a small fraction ever reaching the PSRs (where about one billion tonnes reside). A reasonable estimate of water ever delivered to the lunar surface by comets and asteroids is 0.1–6 trillion tonnes, with perhaps 10–100 billion tonnes not lost to ionization on impact. [6] Interstellar giant molecular clouds might form a significant but not dominant water source, [7] hence at least four basic ways can provide at least 10% of the known PSR water reservoir. Let us consider how each might and how to distinguish among them.

[4] One estimate: the mean OH concentration on the surface is about 100 ppm (Pieters et al. 2009, Clark 2009), applying to about the top 0.5 mm of regolith particles penetrated by 3-micron light. This translates to 150 mg m$^{-2}$, or 5 million tonnes over the entire lunar surface. This is also estimated in terms of the chemical yield of solar wind protons and regolithic oxygen at up to 10–50 million tonnes. ("Sources and physical processes responsible for OH/$H_2O$ in the lunar soil as revealed by the Moon Mineralogy Mapper, M$^3$" by T.B. McCord, L.A. Taylor, J.-P. Combe, G. Kramer, C.M. Pieters, J.M. Sunshine & R.N. Clark, 2011, *Journal of Geophysical Research*, 116, E00G05; see p. 20.) If much of this source were replenished monthly, it could over a few billion years provide hundreds of quadrillions of tonnes, about 1% of the Moon's mass!

[5] "Nominally hydrous magmatism on the Moon" by Francis M. McCubbin, Andrew Steele, Erik H. Hauri, Hanna Nekvasil, Shigeru Yamashita & Russell J. Hemley, 2010, *Proceedings of National Academy of Sciences*, 107, 11223; "Volatile content of lunar volcanic glasses & the presence of water in the Moon's interior" by Alberto E. Saal, Erik H. Hauri, Mauro L. Cascio, James A. Van Orman, Malcolm C. Rutherford & Reid F. Cooper, 2008, *Nature*, 454, 192; "High Pre-Eruptive Water Contents Preserved in Lunar Melt Inclusions" by Erik H. Hauri, Thomas Weinreich, Alberto E. Saal, Malcolm C. Rutherford & James A. Van Orman, 2011, *Science Express*, 10.1126/science.1204626.

[6] "Volatile retention from cometary impacts on the Moon" by Lissa Ong, Erik I. Asphaug, Donald Korycansky & Robert F. Coker, 2010, *Icarus*, 207, 578.

[7] Volatiles from giant molecular clouds (GMCs) might be significant but not dominant: the Moon sweeps up $10^{23}$ cm$^3$ of interstellar volume per second, taking a few million years to transit a GMC. At GMC densities, about $10^3$ atoms cm$^{-3}$, it sweeps up $10^{16}$ moles of gas per GMC. Oxygen composes 1/2000th of all atoms; even if all oxygen swept up is retained as water ice in lunar cold traps, this composes some 100 million tonnes, less than the roughly 1 billion observed. Earth resides in a GMC only about 0.1% of the time, so the Moon has passed through only a few GMCs. Furthermore, for GMCs to deposit any lunar material they must compress the Sun's magnetopause ("heliopause"), where the solar wind arrests oncoming interstellar medium. As seen by *Voyager 1* approaching the heliopause, this is at about 90 A.U. (14 billion km). Within a GMC, the heliopause would collapse in radius as roughly the inverse square-root of interstellar density compared to now, to about 0.01 times its current size. (The Local Interstellar Cloud is 0.1 atom cm$^{-3}$ compared to a GMC's roughly $10^3$ cm$^{-3}$.) The heliopause likely shrinks to Earth's orbit only in the denser parts of GMCs; dense molecular cores, while small, can exceed $10^6$ cm$^{-3}$.

If trillions or quadrillions of tonnes of water formed inside the Moon, can a billion tonnes or more outgas and reach the poles? Water behaves in the magma ocean like an incompatible component, so accumulates in the topmost KREEP layer (*urKREEP*: pristine KREEP after all but about 1% of the magma ocean has crystallized), just below the anorthosite-rich crust, about 60 (±30) kilometers subsurface. [8] This analysis implies average water content cannot be too high, or else concentrations in hydrogen-attracting minerals would be even higher than seen. This seems to betray no more than about 10 ppm water in the magma ocean (still enormous compared to under 1 ppb once discussed). How much oxygen is available to react (its *fugacity*) determines which minerals form from the melt, and may indicate hydrogen playing a role in the lunar mantle as great as water (although hydrogen can form water later).

[8] "Water (hydrogen) in the lunar mantle: Results from petrology and magma ocean modeling" by L.T. Elkins-Tanton & T.L. Grove, 2011, *Earth & Planetary Science Letters*, 307, 173; and "Water in the Lunar Mantle: Results from Magma Ocean Modeling" by L.T. Elkins-Tanton, 2010, *Lunar & Planetary Science Conference*, 41, 1451.

Could water concentrated at the base of the crust have reached the surface? Water-bearing magma could release hydration from the melt by foaming as magma ascended to lower pressures, as volatiles degassed from the urKREEP. Also, rock could be re-heated by later volcanism, as happened in maria, and volatiles would bake out. We should consider more than water; sulfur dioxide composed as much gas in fire fountain glass. The gas once driving fountains might have been carbon monoxide (and some carbon dioxide). [9] There may be signs of this locked in the depths of Cabeus crater: *LCROSS* released a surprising amount of CO, as much as $H_2O$, versus volatiles coming from comets (or solar wind interactions). Comets consist mostly of water, which outweighs CO by 5 – 100 times (and $CO_2$ by a factor of 10 – 50). [10] Most carbon compounds are detected at roughly similar abundances in Cabeus versus comets; CO is an exception. The abundance of CO might betray significant gas from the interior. The chemistry is complicated; this needs further investigation.

[9] "The driving mechanism of lunar pyroclastic eruptions inferred from the oxygen fugacity behavior of *Apollo 17* orange glass" by Motoaki Sato, 1979, *Lunar & Planetary Science Conference*, 10, 311.

[10] "The molecular composition of comets and its interrelation with other small bodies of the Solar System" by Jacques Crovisier, 2006, in *Asteroids, Comets, Meteors 2005: Proceedings of IAU Symposium 229*, eds. D. Lazzaro, S. Ferraz-Mello & J.A. Fernández, Cambridge University Press, p. 133: comets are mostly water, with CO composing only 1-20% of the mass fraction of water, $CO_2$ at 2-10%, $CH_3OH$ at 1-7%, and roughly 1% each for $CH_4$, $H_2CO$, $NH_3$ and $H_2S$.

Atomic argon escapes the lunar interior even now, due to radioactive decay. It finds paths to the surface from deep inside the Moon, and could sweep molecular gases with it. We consider this in Part III.

To reach Cabeus, gas must transit the vacuum. If gas comes from the interior, it must first penetrate the regolith (unless entering the vacuum directly from volcanic vents). Regolith is not always a passive medium. It is nearly impermeable even though made of loose dust, forcing gas particles into circuitous, microscopically tortuous paths as they percolate through it, during which some gases react. A single molecule requires roughly one trillion collisions with particles of regolith before penetrating its entire depth. If the molecule sticks to each regolith particle's surface for even a microscopic time, the molecule's transit requires a macroscopic time interval. If the molecule sticks for a macroscopic time, its transit can take geologically significant times.

Water is special in this regard. Unlike most substances, it can freeze rising through the regolith with the temperatures maintained there. [11] At the equator the temperature just subsurface is 250 K = -23.16ºC, below freezing even at low pressures (over 0.006 atmosphere), and temperatures rise further down (to -10 to 10ºC at the regolith's base [12]) due to heat from the interior. The regolith is colder near the lunar poles. In the presence of water vapor, regolith particles develop a thin water layer. Water is "sticky" in the sense that a water molecule spends a long time stuck to the regolith before freeing itself to travel the next segment of its percolation path, with sticking times highly dependent on temperature. [13] At 0ºC sticking times are only about 1 microsecond; water molecules will percolate out of the regolith within a few days. At regolith temperatures typical of lunar polar regions, around -140ºC, sticking times are about a day, so transit through the regolith takes billions of years. If enough water molecules accumulate stalled in transit, ice will form. Temperatures in PSRs are about -230ºC, so ice can form even near the vacuum.

[11] The organics nonane ($C_9H_{20}$) and benzene ($C_6H_6$) freeze at -13.5ºC and 5.4ºC, respectively, at low pressure, but must be rare in the Moon. Sulfuric acid freezes at 16.8ºC but reacts easily, changing freezing point rapidly with impurities, especially water. (Water's special temperature is unsurprising: Earth is poised near 20ºC, at the same distance from the Sun as the Moon.)

[12] From heat flow experiments on *Apollo 15* and *17* ALSEP ("In-situ measurements of lunar heat flow" by M.G. Langseth & S.J. Keihm, 1977 in *Soviet-American Conference on Geochemistry of the Moon and Planets*, NASA SP-370, p. 283), we know that just below the surface, the stable regolith temperature is 247–253 K (depending on latitude), with gradients (below about 1 m) of 1.2–1.8 deg m$^{-1}$, extrapolating to 0ºC at 13–16 m below the surface, roughly the regolith's depth. A more recent analysis puts the 0ºC level twice as deep, meaning ice could live longer. ("Lost Apollo Heat Flow Data Suggest a Different Lunar Bulk Composition" by Y. Saito, S. Tanaka, J. Takita, K. Horai & A. Hagerman, 2007, *Lunar & Planetary Science Conference*, 38, 2197)

[13] "Subsurface migration of $H_2O$ at lunar cold traps" by Norbert Schorghofer & G. Jeffrey Taylor, 2006, *Journal of Geophysical Research*, 112, E02010.

Ice forming below the Moon's surface, therefore, depends strongly on latitude and how quickly water vapor flows into the regolith from below. Primarily the vapor must reach minimal pressure, 0.006 atmosphere at 0ºC, 0.0001 atmosphere at -50ºC, etc. Given regolith's impermeability, water vapor entering at the base of the regolith at flow rates as low as 0.1 gram per second will form ice (as long as the source is spread out over an area smaller in diameter than the regolith thickness). The incoming flow rate can maintain this pressure over a certain area, which limits how much the ice can grow. An ice patch will add more mass at its edges until so much vapor sublimes from its surface that equilibrium is reached with the incoming flow. At the equator, this area is small (about the size of a football field for 0.1 gram per second) but as one enters the polar regions it increases rapidly, reaching 30 square kilometers at 80º latitude. See Figure 1. [14] The depth where the ice forms probably varies with latitude, 5–10 meters on the equator and more shallow near the poles (depending on the details of regolith properties with depth).

[14] "Lunar Outgassing, Transient Phenomena, and the Return to the Moon. II. Predictions & Tests for Outgassing/Regolith Interactions" by Arlin P.S. Crotts & Cameron Hummels, 2009, *Astrophysical Journal*, 707, 1506.

Water ice in the regolith can change it, however. There is little experimental work on aqueous chemistry in lunar regolith, of course, since there little water was presumed to exist. What is found is that once the oxygen and nitrogen of the air are removed, water eats into regolith grains along damage tracks made by solar wind and cosmic rays. [15] The product of this etching process is calcium pulled from the grains into the interstitial space between them, without changing grain sizes. This new material makes reaching the vacuum even harder for water vapor. Not surprisingly, liquid water is more effective than vapor in this etching, but ice establishes a pseudo-liquid layer on its surface. Indeed lunar re golith chemically and mechanically resembles Portland cement, and tests show that high quality cement can be made from regolith with little processing or additional substances, except sulfur dioxide (to make gypsum). [16]

[15] "Blocking of the water-lunar fines reaction by air and water concentration effects" by R.B. Gammage & H.F. Holmes, 1975, *Proceedings of 6th Lunar Science Conference*, 3, 3305; "Alteration of an annealed and irradiated lunar fines sample by adsorbed water" by H.F. Holmes, P.A. Agron, E. Eichler, E.L. Fuller Jr., G.D. O'Kelley & R.B. Gammage, 1975, *Earth & Planetary Science Letters*, 28, 33.

[16] "Study on Lunar Cement Production Using Hokkaido Anorthite & Hokkaido Space Development Activities" by T. Horiguchi, N. Saeki, T. Yoneda, T. Hoshi, & T. D. Lin, 1996, *Engineering, Construction, and Operations: 5th International Conference on Space*, ASCE Conference Proceedings, 207, 621; and "Behavior of Simulated Lunar Cement Mortar in Vacuum Environment" in *6th International Conference on Space*, 206, 571.

Over geological times water can leach out silicates, which can migrate in solution. [17] One may see calcium hydroxide, magnesium hydroxide, or iron (II) hydroxide ($Fe(OH)_2$). The results are generally alkaline. It is not clear $Fe(OH)_2$ would oxidize to more insoluble rust, $FeO(OH)$. Feldspar e.g., anorthosite is common in many areas, so water might make clays. It seems that sulfur dioxide is released with water, so sulfuric acid might be expected, which would produce gypsum ($CaSO_4 \cdot 2H_2O$) from leached calcium. Theory indicates possibly abundant carbon dioxide, unleashing many products once it forms carbonic acid ($H_2CO_3$): for instance, olivine (($Mg,Fe)_2SiO_4$) or pyroxene (typically $(Ca,Mg,Fe)SiO_3$) can make talc ($Mg_3Si_4O_{10}(OH)_2$), serpentine ($Mg_3Si_2O_5(OH)_4$ plus methane), or it may dissolve, albeit slowly here.

[17] See "Reaction Kinetics of Primary Rock-forming Minerals under Ambient Conditions" by S.L. Brantly, 2004, *Treatise on Geochemistry*, 5, 73.

Hydration-affected minerals have been found, although rarely, mostly from *Apollo 16*, notably more a highlands site. Two meters below the surface in core sample 60002 various oxyhydrates of iron were found. [18] "Rusty Rock" 66095 was chipped from a large boulder excavated only about 1 million years ago but formed about 1 billion years ago. It contains various hydration-affected minerals, particularly goethite ($FeO(OH)$ – also found in some *Apollo 14* samples). Originally terrestrial contamination was suspected, then cometary impact. Evidence against both exists. Many still consider the chemistry of 66095 a mystery, but fumarolic origin seems plausible. [19]

[18] "*Apollo 16* deep drill - A review of the morphological characteristics of oxyhydrates on rusty particle 60002, 108, determined by SEM" by Stephen E. Haggerty, 1978, *Lunar & Planetary Science Conference*, 9, 1861.

[19] "*Apollo 16* 'Rusty Rock' 66095" by L.A. Taylor, H.K. Mao & P.M. Bell, 1973, *Abstracts of Lunar & Planetary Science Conference*, 4, 715; "Origin of inert gases in 'rusty rock' 66095" by D. Heymann & W. Hübner, 1974, *Earth & Planetary Science Letters*, 22, 423; "Chlorine Isotope

Composition of 'Rusty Rock' 66095 & *Apollo 16* Soil. Implications for Volatile Element Behavior on the Moon" by C.K. Shearer & Z.D. Sharp, 2011, *A Wet Vs. Dry Moon*: LPI, 6006.

    This complex chemical system will likely not to be fully understood without simulation experiments at least, but it seems that water's effect on regolith may be primarily to fill in vacuum spaces in the regolith and produce hydrated minerals. More complexity enters considering what these systems do over geological time: the lunar interior is cooling, and the Sun is heating up. [20] These would make the temperature just below the regolith surface much colder, but the temperature gradient steeper, pushing closer to the surface where ice grows. As time proceeds, the ice layer might form deeper into the regolith, leaving water-modified rock in a cap above it. Regolith thickness grows during this time, and impacts will slowly fracture this rock cap. [21] If the water vapor continues to flow, does it leak out, or does it tend to fill in the cracks?

[20] "Solar interior structure & luminosity variations" by D.O. Gough, 1981, *Solar Physics*, 74, 21.

[21] This process likely does not dominate, since even overturn to depths as small as 1 meter takes more than 1 Gigayear ("Mixing of the Lunar Regolith" by D.E. Gault, F. Hoerz, D.E. Brownlee & J.B. Hartung, 1974, *Abstracts of Lunar & Planetary Science Conf.*, 5, 260; "Development of the mare regolith – Some model considerations" by William Quaide & Verne Oberbeck, 1975, *The Moon*, 13, 27). Craters 75 meters in diameter will permanently excavate to 15-meter depth (e.g., Collins 2001, and ignoring effects of crack and breccia formation), and are formed at a rate of about 1 $Gy^{-1}$ per $km^2$ (using a Shoemaker number/size power-law index 2.9 to extend Neukum et al.: "Cratering Records in the Inner Solar System in Relation to the Lunar Reference System" by G. Neukum, B.A. Ivanov & W.K. Hartmann, 2001, *Space Science Reviews*, 96, 55).

    The possibility of hydrated materials concentrated towards the lunar poles was predicted by this model before they were dramatically revealed by *Chandrayaan-1*/$M^3$ in 2009, and it predicts a second effect: especially near the poles, one might expect patches where hydrogen concentration is elevated even far from PSRs. These were discovered in 2010 in the neutron absorption signal. [3] While other models may explain the first pattern, the second may require outgassing from the interior.

How do gas molecules propagate across the lunar surface to the poles? They jump. The Moon has a "ballistic atmosphere" in the sense that atoms and molecules are kicked from the surface by thermal vibrations then enter an elliptical orbit (closely approximated by a parabola) and follow it usually until they land elsewhere on the surface. With the exception of hydrogen and helium, atoms and molecules will bounce many times (or stick) before rebounding fast enough to escape the Moon completely (2.4 kilometers per second or more). Radon-222 is a heavy, radioactive gas that bounces within a small area, random walking only about 100 kilometers before decaying.

How much water propagates to the poles? Ballistic transport models of water molecules tend to find only a small fraction sticking in polar cold traps. [6,22] Butler found 35% of atmospheric water molecules would freeze into PSRs. Stewart et al. find only 1/30th would eventually freeze into PSRs, and those come from the only 3% of water not lost from the Moon immediately in the case of a comet impact. Ong et al. find about 16% and 6.5% for these values, respectively, so about 10 times more delivered to the poles from comets than in Stewart et al. Crider and Vondrak found 4.2% of water molecules in the atmosphere would freeze in PSRs. They studied water created by chemical reaction of hydrogen in the solar wind reacting with the oxygen-rich lunar regolith. For other molecules, we can expect to find substances of masses about 10 to 100 a.m.u.: less than this, and molecules achieve escape velocity on the lunar day side, more than this and their random walk rarely reaches the polar regions.

[22] "The migration of volatiles on the surfaces of Mercury & the Moon" by Bryan J. Butler, 1997, *Journal of Geophysical Research*, 102, 19283; "The solar wind as a possible source of lunar polar hydrogen deposits" by Dana Hurley Crider & Richard R. Vondrak, 2000, *Journal of Geophysical Research*, 105, 26773; "Simulations of a comet impact on the Moon & associated ice deposition in polar cold traps" by Bénédicte D. Stewart, Elisabetta Pierazzo, David B. Goldstein, Philip L. Varghese & Laurence M. Trafton, 2011, *Icarus*, 215, 1. Early models are found in "Orbital Search Lunar Volcanism" by R.R. Hodges, J.H. Hoffman, T.T.J. Yeh & G.K. Chang, 1972, *Journal of Geophysical Research*, 77, 4079.

With the surficial regolith hydroxyl result from M$^3$ in 2009, scientists realize that we should be looking at the production, transport and sticking of OH as much as H$_2$O, but there is significant work to be done. The grains of regolith are covered by a roughly 100-nanometer thickness or "rim" of non-crystalline, amorphous material, in many cases studded with refined-iron inclusions and covered by vapor-deposited volatiles. Solar wind protons typically penetrate of order 10 nanometers, so this rim layer is where they the reaction tends to occur and the barrier to be overcome by escaping products. Production of OH or H$_2$O has been hypothesized for decades, [23] but recent analyses may have detailed more of how hydroxyl is produced by the solar wind. [4] Nonetheless, the flux of solar wind onto the surface is smallest at the poles, but the surficial hydroxyl concentration is greatest there, despite surface temperatures above 0ºC during the day (more than 15 degrees from the poles). How solar wind hydroxyl could produce the observed concentration toward the poles has yet to be worked out.

[23] For instance, "Proton-induced hydroxyl formation on the lunar surface" by E.J. Zeller, L.B. Ronca & P.W. Levy, 1966, *Journal of Geophysical Research*, 71, 4855; "Water detection on atmosphereless celestial bodies: Alternative explanations of the observations" by L. Starukhina, 2001, *Journal of Geophysical Research*, 106, 14701.

The hydroxyl signal across the Moon's surface was observed not only by *Chandrayaan-1*/M$^3$, but also infrared spectrometers on *Deep Impact* and *Cassini*. [24] These spacecraft delivered a hint that the hydroxyl concentration over the course of a lunar day, so surface temperature appears to matter. Furthermore, at a given time of day different minerals will generate or maintain hydration by widely varying amounts: maria less so than highlands, more so for some fresh craters (but not others). Areas around Mare Orientale appear water and hydroxyl rich; the South Pole-Aiken Basin appears water and hydroxyl poor. [25] These differences are not well-understood.

[24] "Detection of Adsorbed Water and Hydroxyl on the Moon" by Roger N. Clark, 2009, *Science*, 326, 562; "Temporal and Spatial Variability of Lunar Hydration As Observed by the Deep Impact Spacecraft" by Jessica M. Sunshine, Tony L. Farnham, Lori M. Feaga, Olivier Groussin, Frédéric Merlin, Ralph E. Milliken and Michael F. A'Hearn, *Science*, 326, 565.

[25] "Water & Hydroxyl on the Moon as Seen by the Moon Mineralogy Mapper (M$^3$)" by R. Clark, et al., 2010, Lunar & Planetary Science Conference, 41, 2302; also 2011, *A Wet vs. Dry Moon: Exploring Volatile Reservoirs & Implications for Evolution of Moon & Future Exploration*, 6047.

As intriguing as hydration of cometary/asteroidal or solar wind origin may be, the more scientifically interesting water reservoir is inside the Moon, unanticipated by lunar scientists until a few years ago. Theorists show water acted as an incompatible substance in the ancient lunar magma ocean, concentrating near the base of the crust, about 60 kilometers below the surface. [8] This might apply to high concentrations of hydroxyl in apatite, which tends to form at shallow depths. In the case of picritic fire fountain glasses and inclusions within them, an origin much deeper, by at least 400 kilometers, is indicated. [26] On the other hand, the element europium, which can be act like calcium in plagioclase (which floats to the surface early in the magma ocean) is relatively depleted in fire fountain glasses. [27] This argues that the magma that produced these glasses arose in the magma ocean below the crust (but above the earliest, deepest layer which is olivine-rich).

[26] "Magmatic processes that produced lunar fire fountains" by Linda T. Elkins-Tanton, Nilanjan Chatterjee & Timothy L. Grove, 2003, *Geophysical Research Letters*, 30, 1513.

[27] "Basaltic magmatism on the Moon: A perspective from volcanic picritic glass beads" by C.K. Shearer & J.J. Papike, 1993, Geochimica et Cosmochimica Acta, 57, 4785.

The search is on for caveats that would explain the presence of water in the Moon without destroying the giant impact hypothesis of lunar origin. The models of water's role in the magma ocean supports the notion that the Moon still is largely depleted, to perhaps 10 ppm compared to several hundred ppm for Earth, and this is a level of that can be made consistent with the Giant Impact model. As material fell back to Earth and congealed into the Moon, it formed a proto-lunar disk made primarily of material from the extraterrestrial impactor. While this material was very hot and lost most of its volatiles, some remained in the environment and sufficient amount remained entrained in order to support a magma ocean bulk composition of 10 ppm water. [28]

[28] "A Model of the Moon's Volatile Depletion" By Steven J. Desch & G. Jeff Taylor, *A Wet vs. Dry Moon: Exploring Volatile Reservoirs & Implications for Evolution of Moon & Future Exploration*, 6046.

Not all material need to have been processed through the violently heated material of the Giant Impact. If the proto-Earth or the Mars-sized impactor (Theia) had its own satellite, the chances are that it became incorporated into the Moon without being subjected to temperatures sufficient to cook hydration out of rock (about 1200K). Such a satellite of proto-Earth would likely remain in orbit around Earth, and satellites of Theia (which enters at proto-Earth's escape velocity) have about a 50% probability of going into Earth orbit (depending on whether their velocity is directed toward or counter Theia's). A previous satellite might well form the core around which the Moon accumulates, and would be heated to the temperature of the overlying material to eventually but not catastrophically liberate its volatiles.

The prospect of water deep within the Moon is intriguing, because it might help explain the preponderance of deep quakes within the Moon, which on Earth are thought to originate with phase transitions brought on by the presence of water. [29] Deep moonquakes arise in clusters 700 to 1200 kilometers deep, well below the depth of the usually hypothesized magma ocean. [39] There is reason to consider to interior water reservoirs, one consolidated out of the relatively dry magma ocean, and another from the region of the Moon unaffected by the magma ocean, which may or may not be relatively dry.

[29] "The physical mechanisms of deep moonquakes & intermediate-depth earthquakes: How similar & how different?" by C. Frohlich & Y. Nakamura, 2009, *Physics of Earth & Planetary Interiors*, 173, 365.

[30] "Seismic Detection of the Lunar Core" by Renee C. Weber, Pei-Ying Lin, Edward J. Garnero, Quentin Williams & Philippe Lognonné, 2011, *Science*, 331, 309.

Much of our knowledge about the behavior of hydrogen and water in the lunar interior is summarized by the oxygen fugacity $f_{O_2}$, which is the effective pressure of oxygen as if it were an ideal gas, and is related (exponentially) to the chemical potential. A high oxygen fugacity implies a highly oxidized system. Oxygen fugacity can be measured by the equilibrium of oxides of iron, silicon and/or titanium e.g., the dissociation of ilmenite ($FeTiO_3$). Measurements of $f_{O_2}$ for lunar samples tend to imply mantle values several orders of magnitude below those for Earth's mantle. On Earth this implies that most mantle hydrogen is oxidized to water, whereas on the Moon one should expect much more $H_2$. This has implications, given that hydrogen diffuses much more quickly through mineral lattices than water, predicting rapid loss of hydrogen. With the loss of hydrogen, at high temperature water would tend to dissociate (primarily by oxidizing iron). If magma is quenched rapidly, this might not occur. This might explain the hydrated glass samples, but these lunar fire fountain glasses show the highest $f_{O_2}$ values of any lunar samples. [9] Elkins-Tanton & Grove [8] argue that hydrogen must be an important, perhaps dominant, species in partition with water, but still allow for significant water. [31]

[31] The water/hydrogen system is complicated in silicate melts by the dissolution of hydrogen as hydroxyl, significantly reducing the concentration of hydrogen relative to water. The rapid diffusion of hydrogen means that it outgasses quickly. (See "Formation of carbon & hydrogen species in magmas at low oxygen fugacity" by A. Kadik, F. Pineau, Y. Litvin, N. Jendrzejewski, I., Martinez & M. Javoy, 2004, *Journal of Petrology*, 45, 1297; "Influence of Oxygen Fugacity on the Solubility of Carbon & Hydrogen in $FeO-Na_2O-SiO_2-Al_2O_3$ Melts in Equilibrium with Liquid Iron at 1.5 GPa & 1400°C" by A.A. Kadik, N.A. Kurovskaya, Y.A., Ignat'ev, N.N. Kononkova, V.V. Koltashev &V.G. Plotnichenko, V.G., 2010, *Geochemistry International*, 48, 953; "Nitrogen & hydrogen isotope compositions & solubility in silicate melts in equilibrium with reduced (N plus H)-bearing fluids at high pressure & temperature: Effects of melt structure" by B.O. Mysen & M.L. Fogel, 2010, *American Mineralogist*, 95, 987; "'Water' in lunar basalts: The role of molecular hydrogen ($H_2$), especially in the diffusion of the H component" by Y. Zhang, 2011. *Lunar & Planetary Science Conference*, 42. 1957; "Diffusion of H, C & O components in silicate melts" by Y. Zhang & H. Ni, 2010, in *Diffusion in Minerals & Melts,* eds. Y. Zhang & D.J. Cherniak, Chantilly, VA: Mineralogical Society of America, pp. 171-225.)

Sharp et al. [32] argue that the importance of water in maintaining the constant $^{35}Cl/^{37}Cl$ ratio on Earth and its heterogeneity on the Moon implies that water is absent from the lunar interior. The low lunar oxygen fugacity encourages decoupling of H and Cl degassing, however, as demonstrated experimentally. [33] Alternatively, this might indicate evidence of large inhomogeneity, both as a function of depth and different locations across the surface.  Perhaps the Moon is more complex than we imagine.

[32] "The Chlorine Isotope Composition of the Moon & Implications for an Anhydrous Mantle" by Z.D. Sharp, C.K. Shearer, K.D. McKeegan, J.D. Barnes & Y.Q. Wang, 2010, *Science*, 329, 1050.

[33] "Differential degassing of H$_2$O, Cl, F & S: Potential effects on lunar apatite" by Gokce Ustunisik, Hanna Nekvasil & Donald Lindsley, 2011, *American Mineralogist*, October, 96, 1650.

There is much that the model of vacuum deposition into polar cold trap model that does not explain.  While it is true that some of the strongest signals of hydrogen concentration measured via radar circular polarization return or epithermal neutron absorption are confined to PSRs Cabeus (diameter = 101 km, Latitude = -85.3°, depth = 4.5 km) and Shoemaker (48 km, -88.0°, 4 km), while equally favorable PSRs Faustini (42 km, -87.18°, 3.5 km) and Haworth (51 km, -87.2°, 3.5 km), show no such signal.  Furthermore, several patches which are not permanently shadowed but 10 degrees or more from the poles nonetheless show strong epithermal neutron absorption signals.  Some other factors determine the emplacement of hydrogen as water or otherwise.  If one accepts that outgassing 3 – 4 billion years ago included water that might stick within the regolith for geologically long times, or that more recent outgassing might contain water vapor, this might explain this variation in hydration signal across the polar regions.  The origin of water in PSRs is amenable to isotopic analysis.  Solar wind hydrogen is low in deuterium.  Endogenous lunar oxygen has oxygen isotope abundance ratios similar to Earth's.  This should be a relatively straightforward signature for surficially-generated hydroxyl.  The deuterium enhancement $\delta D/1000 = [(D/H)_{sample}/(D/H)_{ocean}] - 1$, where "ocean" refers to the Vienna standard mean ocean. Terrestrial water samples range in $\delta D$ from -500‰ to +100‰, with most values closer to zero by at least a factor of two.  Comets have $\delta D$ values that tend to cluster around +1000‰ or larger, whereas chondritic meteorites vary over a wide range.  The interstellar D/H ratio varies between molecules and states of matter in complex ways, however. [34] Meteoritic oxygen isotope ratios tend to cluster close and symmetrically around lunar/terrestrial values, however. [35] In contrast, lunar apatite δD was found to vary from +391‰ to +1010‰, with highlands apatite having +240‰ +340‰, and some mare basalts as low as -172‰ to -215‰. [35]

[34] "Oxygen Isotopes in Meteorites" by R.N. Clayton, 2003, *Treatise on Geochemistry*, 1, 129.

[35] "Interstellar Ices as Witnesses of Star Formation: Selective Deuteration of Water & Organic Molecules Unveiled" by S. Cazaux, P. Caselli & M. Spaans, 2011, *Astrophysical Journal Letters*, 741, L34.

[36] "Hydrogen isotope ratios in lunar rocks indicate delivery of cometary water to the Moon" by James P. Greenwood, Shoichi Itoh, Naoya Sakamoto, Paul Warren, Lawrence Taylor & Hisayoshi Yurimoto, 2011, *Nature Geoscience*, 4, 79.

One pictures sampling water in PSRs by robotic mission, but that the signals from possible sources, meteoritic, cometary, solar wind interactions, endogenous or interstellar being mixed by regolith impact gardening in a way destructive of the separation of information regarding these different contributions. This does not need to be the case. Imagine topographic features that are dramatically higher or lower than the surrounding terrain. A positive elevation projection will be worn down at a roughly linear rate (modulo stochastic variation in impacts and longterm evolution in average impact rate over Solar System history). Likewise, a deep depression will be filled in by ejecta from adjacent impacts at a roughly linear rate. In such regions a roughly time-ordered stratigraphy should allow a monotonic sequence of events to be uncovered. The ratio of this process to more dispersive mixing by gardening depends on the local topography, and we are constructing a tool to allow one to estimate the relative importance of these processes, given a digital elevation model. [37] Episodes of asteroidal or cometary impact, incursion of dense molecular interstellar cores into the inner Solar System, episodes of anomalous solar wind flux, or endogenous outgassing events may be recorded within the vertical structure of these regions. Such a tool might tell us where best to drill. Water from these various sources will be layered with different telltale substances ($SO_2$ and $CO/CO_2$ from outgassing, different organics from comets/asteroids versus interstellar clouds). Together these time-ordered compositional and isotopic clues should suffice to describe what sources contributed to the lunar atmosphere, when and how much.

[37] "Gardening versus Secular Deposition of Regolith: A Numerical Model" by Andrew Hodges & Arlin P.S. Crotts, 2012, in preparation.

    Robotic sampling and drilling in PSRs is far from the only approach to settling many of these issues, however. For instance, we still have not confirmed the hydration signal reported in 2-meter core samples from *Luna 24*, [38] which could be explored now with infrared spectroscopy, pyrolysis and/or secondary ion mass spectrometry. Landers are not the only mission architecture that could shed significant light on the nature of lunar volatiles, especially water. The interaction of water with the lunar surface is inherently an issue of the regolith, and current probes only extend about a meter or less into the lunar soil. Ground penetrating radar of about 400 Megahertz could image the entire depth of the regolith at better than 1 meter resolution, and would easily unearth the signatures of subsurface ice or volatile-altered regolith. While *Kaguya* carried an alpha particle spectrometer, sensitive to outgassing radon-222 and its daughter polonium-210, this instrument's power bus suffered a catastrophic failure and thus far no useful data have been published. While the *Lunar and Dust Environment Explorer* (*LADEE*) will study constituents of the lunar atmosphere, it will only do so in a narrow strip along the lunar equator. To understand how hydroxyl and other species migrate to the poles, or to find localized sources of lunar outgassing, a mass spectrometer orbiting the Moon from pole to pole is required. A followup instrument completing the work of $M^3$ could finish mapping the hydroxyl distribution across the Moon, study how it changes over the course of a month, and extend the wavelength range sufficiently around 3 microns to unambiguous separate the absorption signature from hydroxyl versus water in mineral matrices (and versus water ice). One could imagine a more sensitive and higher resolution map of epithermal neutron absorption, as well. These last five instruments could easily share the same lunar polar orbiter, along with instruments for other purposes.

[38] "Possible Water in Luna 24 Regolith from the Sea of Crises" by M.V. Akhmanova, B.V. Dementyev & M.N. Markov, 1979, *Geochemistry International*, 15, 166.

   Much of the excitement surrounding lunar volatiles arises not just from the science, but from their potential practical utilization.  We know there is a high concentration in the upper meter or more of regolith within PSRs, and suspect there is an abundance of perhaps 1 part per thousand in the upper millimeter or so in the larger polar regions.  There may be more: if outgassing from the interior into the regolith of the polar regions is occurring or has occurred, there may be an even larger, more concentrated resource a few meters down.  The potential is that these sites may be at temperatures of 150K or 200K, more like Antarctic temperatures rather than near absolute zero temperatures (20-50K) in the PSRs.  We may be discussing solid ice bodies, unlike the surficial signal evident at 3 microns.  Scenarios have been cast for exploitation of the PSR or surficial hydration resource, a subsurface regolith reservoir might be even easier to exploit.

   Why be excited about lunar hydration resources?  There are several reasons, both obvious and not so obvious.  With current transportation costs of order $100,000 per kilogram, a source of water so necessary for human life is a crucial asset.  Furthermore, water is propellant.  With abundant electricity derived from solar energy, water provides liquid oxygen and liquid hydrogen, in many ways an optimal rocket bipropellant.  Even if water is not mined from the PSRs, the fact that cold storage of liquid oxygen and hydrogen (with critical temperatures of 154.6K and 33.2K, respectively, warmer than some PSRs) in the polar regions will be relatively easy makes their use in rocket applications more practicable.  Furthermore, liquid oxygen and hydrogen make an easy "battery" for energy storage and electricity production in fuel cells.  Furthermore, water is very useful in conjunction with lunar regolith, which has chemical and mechanical properties e.g., particle size distribution, usefully similar to cement, such that studies indicate it might make a completely indigenous and adaptable building material.  Also, water provides excellent shielding for high-energy protons, especially from the solar wind, and spallation products e.g., neutrons.

Real efforts are underway to utilize lunar water as an exploration resource, particularly commercially.  Several issues should be clarified.  As mentioned the site of the most accessible lunar water is unclear.  Water detected by infrared absorption is easily accessible but at low concentrations, at the level of a few kilograms per hectare.  Much more concentrated sources (hundreds or thousands of tonnes per hectare) are available in the PSRs, but only for machinery that can operate within 50 degrees of absolute zero.  The possibility of water frozen into the regolith due to outgassing needs to be explored.  Space law regarding the utilization of lunar water is unclear.  It may require a test case before the United Nations or other body to establish which uses are legal.  Some hold that substances trapped with the PSRs should be serve as common legacy to all of humanity, and should not be subject to exploitation by any one country or commercial concern.  Even without this impediment, water from PSRs is contaminated with poisons of an isotopic (deuterium fraction) and chemical nature (mercury and other heavy metals, and perhaps organic poisons).  Some of the organic compounds in PSRs are potentially valuable in their own right, since carbon and nitrogen are otherwise extremely rare on the lunar surface.  Even without exploitation of substantial resources, however, the conditions within PSRs by themselves can be an asset.  The PSRs are some of the coldest and most stable locations in the Solar System, and objects kept there might be expected to remain unaltered on the scale of Gigayears.  Whether it is the cultural or genetic legacy of humanity, or the individual remains of the superwealthy seeking their own pharoanic vision of eternity, we can store things at the lunar poles with a security heretofore unimagined.

The availability of abundant liquid oxygen and hydrogen could be a boon in many ways.  Given this propellant source, the easiest way to imagine transportation on the Moon beyond the range of tens of kilometers is via a ballistic "hopper" which accelerates and decelerates itself and payload to/from ballistic arcs of one hour's duration or less.  A moon base at the North and/or South Pole could access the surface with robotic or human-carrying expeditions.  Since access to free space from the lunar surface is roughly 30 times easier than from Earth, the Moon could act as a supply depot for missions preparing to accelerate towards more distant reaches of the Solar System, with their parting impulse derived primarily from chemical rockets fueled from the Moon. One can imagine an industry of spacecraft using lunar propellant transiting cislunar space in order to service satellites in geosynchronous orbit, which otherwise would be lost as they deplete fuel and drift out of useful orbits for communications or remote sensing. A "space tug" using lunar liquid oxygen and hydrogen could restore them to their operational orbits and extend their lifetimes by many-fold, then escape back to lunar vicinity for propellant recharge and a new cycle across cislunar space. [39]

[39] We outline technical details of such a system in a pending patent documents, to appear later.

The impressive progress made in understanding volatiles such as water is the product primarily of new technology, and to some extent the fit of earlier data into a developing paradigm that expands on the profound discoveries of Apollo and other earlier missions. There are further phenomena associated with volatiles in the lunar environment that are sit further beyond the periphery of the discovery process that might lead to further insights into the lunar interior and evolution, and these we treat in Part III.

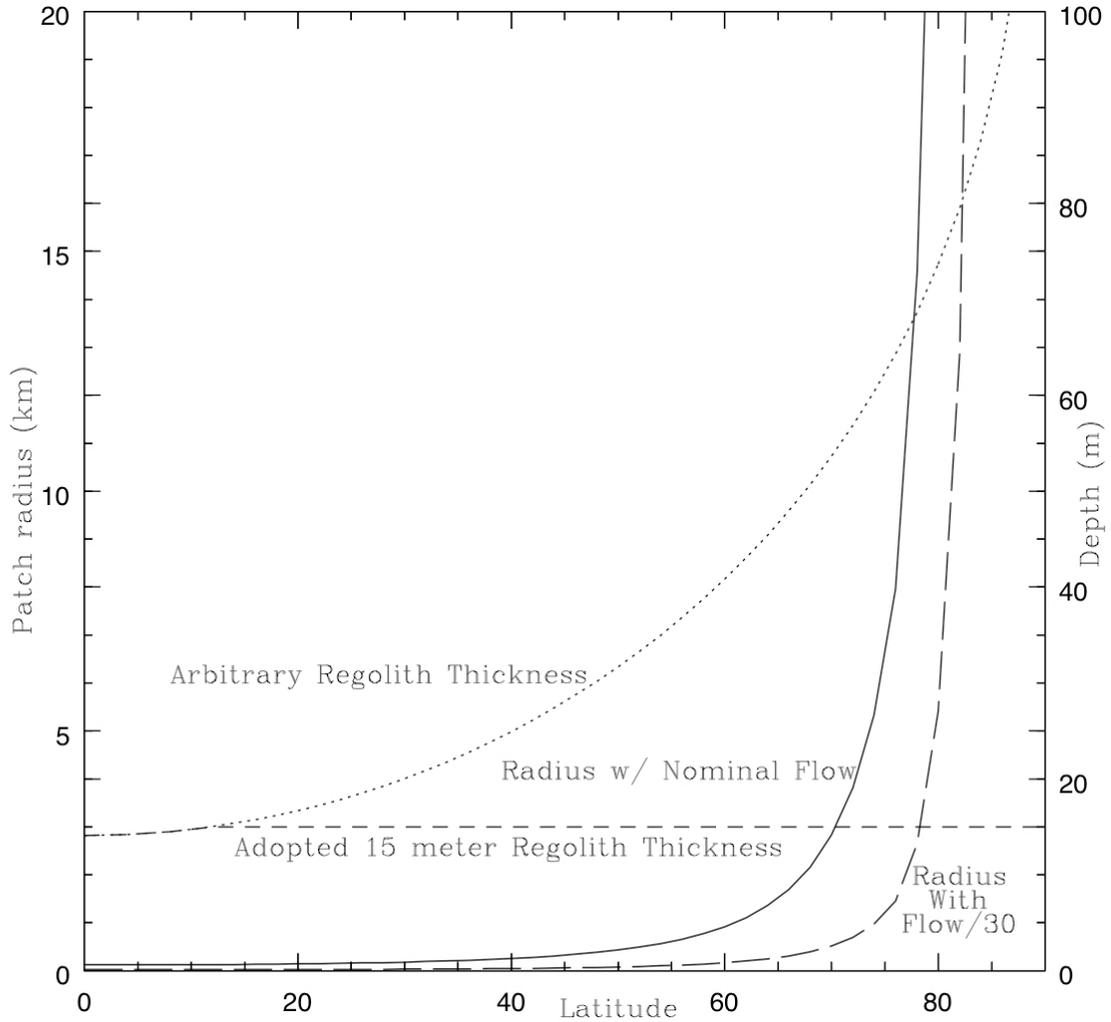

**Figure 1:** Tendency of seeping water vapor to form subsurface ice, vs. lunar latitude (from Crotts & Hummels 2009). For point-source seepage, the radius of ice patch formed is shown by the solid curve (use left axis) and the preferred depth by short dashed curve (use right axis). The arbitrary 15 meter limit is adopted assuming that the low diffusivity regolith overlays a higher diffusivity megaregolith discouraging ice growth. If regolith is actually deeper, or if an ice cap might actually encourage ice growth at greater depth, the ice layer might extend to dotted curve (reading right axis), this would likely encourage more ice growth at given flow rate. (If regolith were surprisingly deep, ice patch area might grow larger by a factor roughly the ratio of the dotted curve to the dashed curve.) The long-dashed curve is similar to the solid curve, showing the ice patch size if the flow rate is reduced by a factor of 30 (to 0.0033 g s$^{-1}$ of water). This curve does not account for time required to reach equilibrium radius. Smaller than this flow, ice patches might not grow near the equator.